\documentclass[10pt]{article}
\usepackage{amsfonts}
\usepackage{amsmath}
\title{Non-relativistic Bremsstrahlung in QED: \\Hamiltonian vs. Path Integral Approach}
\author{A. Jahan\\Research Institute for Astronomy and Astrophysics of Maragha
(RIAAM)\\ Maragha, IRAN, P. O. Box: 55134 - 441\\jahan@riaam.ac.ir}
\date{}
\usepackage[english]{babel}
\begin{document}
\maketitle
\begin{abstract}
An alternative derivation of the radiation intensity in non-relativistic bremsstrahlung is provided utilizing the path integral formalism. By integrating out the gauge field, one obtains the effective action which it's imaginary part is interpreted as the rate of photon production during the collision.
\end{abstract}
\section{Introduction}
The non-relativistic bremsstrahlung is among those subjects discussed by few quantum field theory textbooks [1-3]. The bremsstrahlung cross section was first calculated by A. Sommerfeld [4]. Such a tedious calculation involves handling the exact wave function of the attractive Coulomb field in terms of the confluent hypergeometric functions within the framework of first order perturbation theory. The result obtained using the exact wave function coincides with one obtained using the Born approximation if for a particle with incoming and outgoing momenta $\textbf{p}=m\textbf{v}$ and $\textbf{p}'=m\textbf{v}'$ the condition
\begin{equation}\label{1}
Ze^2\ll v,v'.
\end{equation}
holds [1, 2]. Having at hand the cross section, one can calculate the intensity of emitted radiation.\\
In this note, we give another formulation of the problem using the path integral formalism. We consider a non-relativistic spinless particle interacting with a quantized electromagnetic field and derive the radiation intensity in a scattering from the Coulomb potential caused by a point-like source with charge $-Ze$. We follow the trick of [5-7] to integrate out the electromagnetic field, which gives rise to the effective action $S_{\textrm{eff}}$. The imaginary part of $S_{\textrm{eff}}$ yields the probability of photon emission during the collision. The real part of the effective action is responsible for the energy shift of atomic levels [5, 6]. In next section, we review the approach based on the first order perturbation theory to deal with the non-relativistic bremsstrahlung. In section 3, we engage the machinery of path integrals to re-derive the radiation intensity in a non-relativistic collision. The goal of this work is to demonstrate that how the effective action formalism as discussed above can provide us with the radiation intensity of photon emission in a non-relativistic scattering. In this work we use the natural units $\hbar=c=1$. Also summation over the repeated indices is assumed.
\section{Non-relativistic Bremsstrahlung}
The interaction Hamiltonian between a non-relativistic particle and photon field could be obtained by minimal substitution $\frac{\widehat{\textbf{p}}^2}{2m}\rightarrow\frac{1}{2m}{(\widehat{\textbf{p}}-e\textbf{A})^2}$ yielding
\begin{equation}\label{1}
H_{\scriptsize\textrm{int}}=-\frac{e}{m}\widehat{\textbf{p}}\cdot \textbf{A}.
\end{equation}
The Fourier expansion of the vector potential in terms of the creation and destruction operators in the transverse gauge $\nabla\cdot \textbf{A}=0$, in a unite volume ($V=1$) is
\begin{equation}\label{1}
\textbf{A}=\sum_{\lambda=1}^2\sum_{\scriptsize\textbf{k}}\sqrt{\frac{2\pi}{\omega_k}}\Big({{\epsilon}_{\lambda\scriptsize\textbf{k}}}
a_{\lambda\scriptsize\textbf{k}}e^{-i(\omega_k t-\scriptsize{\textbf{k}\cdot \textbf{x}})}+{\overline{\epsilon}_{\lambda\scriptsize\textbf{k}}}{a}^{\dagger}_{\lambda\scriptsize\textbf{k}}e^{i(\omega_k t-\scriptsize{\textbf{k}\cdot \textbf{x}})}\Big),\qquad \omega_k=|\textbf{k}|=k
\end{equation}
such that ${\epsilon}_{\lambda\scriptsize\textbf{k}}\cdot \textbf{k}=0$. We assume an incoming particle with momentum $\textbf{p}$ and no photon as the initial state  $|i\rangle=|\textbf{p},0\rangle$, while the final state  $|f\rangle=|\textbf{p}',\textbf{k}\rangle$ assumed to be an outgoing particle with momentum $\textbf{p}'$ together with an emitted photon with momentum $\textbf{k}$. The particle's wave function assumed to satisfy
\begin{equation}\label{1}
H\phi_{\scriptsize\textbf{p}}(\textbf{x})=\bigg(-\frac{1}{2m}\nabla^2-\frac{Ze^2}{2|\textbf{x}|^2}\bigg)\phi_{\scriptsize\textbf{p}}(\textbf{x})=E_p\phi_{\scriptsize\textbf{p}}(\textbf{x})
\end{equation}
The amplitude for a transition between the initial and final states in dipole approximation, i.e. $e^{-i\scriptsize{\textbf{k}\cdot \textbf{x}}}\simeq 1$, becomes
\begin{equation}\label{1}
\mathcal A_{fi}=\langle f|H_{\scriptsize\textrm{int}}|i\rangle=-\overline{\epsilon}_{\lambda\scriptsize\textbf{k}}\cdot\widehat{\textbf{p}}_{\scriptsize{\textbf{p}'\textbf{p}}}
\frac{e}{m}\sqrt{\frac{2\pi}{\omega_k}}.
\end{equation}
where $\widehat{\textbf{p}}_{\textbf{p}'\textbf{p}}\equiv\langle\textbf{p}'|\widehat{\textbf{p}}|\textbf{p}\rangle$. So, the differential cross section for the bremsstrahlung scattering is found to be
{\setlength\arraycolsep{2pt}
\begin{eqnarray}\label{1}
d\sigma_{\lambda\scriptsize{\textbf{kp}'}}&=&\frac{2\pi}{v}\frac{|\mathcal A_{fi}|^2}{T}\frac{d^3k}{(2\pi)^3}\frac{d^3p'}{(2\pi)^3}\delta(E_p-E_{p'}-\omega_k),\\\nonumber
&=&\frac{4\pi^2e^2}{vm^2}|\overline{\epsilon}_{\lambda\scriptsize\textbf{k}}\cdot
\widehat{\textbf{p}}_{\scriptsize{\textbf{p}'\textbf{p}}}|^2\frac{d^3p'}{(2\pi)^3}\frac{d^3k}{(2\pi)^3k}\delta(E_p-E_{p'}-k).
\end{eqnarray}}
where $v=\frac{\textbf{p}}{m}$ and $\omega_{pp'}=E_p-E_{p'}$ denote the incoming flux and photon energy respectively. Therefore, the radiated power will be
{\setlength\arraycolsep{2pt}
\begin{eqnarray}\label{1}
P=\sum_{\lambda=1}^2\int d\sigma_{\lambda\scriptsize{\textbf{kp}'}}\omega_{pp'}&=&\frac{4\pi^2e^2}{vm^2}\sum_{\lambda=1}^2\int\frac{d^3p'}
{(2\pi)^3}\int\frac{d^3k}{(2\pi)^3k}\,\omega_{pp'}|\overline{\epsilon}^
\lambda_{\scriptsize\textbf{k}}\cdot\widehat{\textbf{p}}_{\scriptsize{\textbf{p}'\textbf{p}}}|^2\delta(E_p-E_{p'}-k),\\\nonumber
&=&\frac{4e^2}{3}\int\frac{d^3p'}{(2\pi)^3v}\,\omega_{pp'}^4|\textbf{x}_{\scriptsize{\textbf{p}'\textbf{p}}}|^2,\\\nonumber
&=&\frac{4e^2}{3}\int\frac{d^3p'}{(2\pi)^3v}\,|\ddot{\textbf{x}}_{\scriptsize{\textbf{p}'\textbf{p}}}|^2,
\end{eqnarray}}
where we have invoked $\widehat{\textbf{p}}_{\scriptsize{\textbf{p}'\textbf{p}}}=m\dot{\textbf{x}}_{\scriptsize{\textbf{p}'\textbf{p}}}=im\langle \textbf{p}'|[H,\textbf{x}]| \textbf{p}\rangle=im\omega_{pp'}\textbf{x}_{\scriptsize{\textbf{p}'\textbf{p}}}$ and
\begin{equation}\label{1}
\int d\Omega_k P_{ij,\scriptsize\textbf{k}}=\int d\Omega_k
\sum_{\lambda=1}^2\epsilon_{i,\lambda\scriptsize\textbf{k}}\overline{\epsilon}_{j,\lambda\scriptsize\textbf{k}}=\frac{8\pi}{3}\delta_{ij}.
\end{equation}
Thus, with the aid of classical equation of motion
\begin{equation}\label{1}
m\ddot{\textbf{x}}=\frac{Ze^2}{2}\nabla\frac{1}{|\bf x|},
\end{equation}
we find
\begin{equation}\label{1}
P=\frac{4e^2}{3}\Big(\frac{Ze^2}{2m}\Big)^2\int\frac{d^3p'}{(2\pi)^3v}\,|\langle\textbf{p}'|\nabla\frac{1}{|\bf x|}|\textbf{p}\rangle|^2
\end{equation}
Now, let us look at the classical problem of radiation. By assuming a Fourier expansion for the particle's position as
\begin{equation}\label{1}
\textbf{x}(t)=\int_{-\infty}^\infty \frac{d\omega}{2\pi}\,\textbf{x}(\omega)e^{i\omega t},
\end{equation}
the time-averaged intensity reads
{\setlength\arraycolsep{2pt}
\begin{eqnarray}\label{1}
\langle P \,\rangle=\frac{1}{T}\int_0^Tdt\,P(t)&=&\frac{2e^2}{3}\frac{1}{T}\int_0^Tdt\,\ddot{\textbf{x}}^2(t),\\\nonumber
&=&\frac{4e^2}{3}\int_{0}^\infty d\omega\,\omega^4 |\textbf{x}(\omega)|^2,\\\nonumber
&=&\frac{4e^2}{3}\int_{0}^\infty d\omega\,|\ddot{\textbf{x}}(\omega)|^2,\\\nonumber
&=&\int_{0}^\infty d\omega\,P(\omega).
\end{eqnarray}}
Therefore, we observe that the quantum mechanical radiation power formula (7) in a collision process is attainable form the Fourier transform of the classical formula on substituting
\begin{equation}\label{1}
P(\omega)=\frac{4e^2}{3}|\ddot{\textbf{x}}(\omega)|^2\longrightarrow P_{\scriptsize{\textbf{p}'\textbf{p}}}=\frac{4e^2}{3}|\ddot{\textbf{x}}_{\scriptsize{\textbf{p}'\textbf{p}}}|^2.
\end{equation}
The above correspondence between the Fourier component of the classical system and the transition matrix elements is a typical example of the so called "Quasi-classical" behavior. The radiation is quasi-classical when it's energy is small in comparison with the initial and final energies of the particle [1]. So, starting from the classical radiation intensity and taking into account the quantum-classical correspondence (13) one can obtain the radiation intensity (7) via multiplying the term $\frac{4e^2}{3}|\ddot{\textbf{x}}_{\scriptsize{\textbf{p}'\textbf{p}}}|^2$ by the factor $\frac{d^3p'}{(2\pi)^3v}$ and integrating over the final states.\\
As the final remark, let us remind a simple fact from the classical electrodynamics: a freely moving charge does not radiate. Quantum mechanically, this means that for a free charged particle the first order perturbation theory makes no sense and one must embark on the second order perturbation to take into account the effect of Coulomb potential [3]. However, the legitimacy of the first order perturbation theory gained to derive (7), refers to the fact that the effect of Coulomb potential is already encoded in the particle's (exact) wave function.
\section{Path Integral Approch}
The Lagrangian of a non-relativistic particle interacting with the gauge field in transverse (Coulomb) gauge is
\begin{equation}\label{1}
L=\frac{m}{2}\dot{\textbf{x}}^2+\frac{Ze^2}{2|\textbf{x}|}+e\int d^3x\,\dot{\textbf{x}}\cdot \textbf{A}+\frac{1}{8\pi}\int \textbf{E}^2_{\perp}-\textbf{B}^2d^3x.
\end{equation}
where $\textbf{B}=\nabla\times \textbf{A}$ and $\textbf{E}_{\perp}=-\dot{\textbf{A}}$. We rewrite the last two terms of (14) as
\begin{equation}\label{1}
L_{\scriptsize\textrm{rad}}+L_{\scriptsize\textrm{int}}=\frac{1}{8\pi}\int d^3x A^iD^{-1}_{ij}A^j +e\int d^3x\,\dot{x}_i A^i.
\end{equation}
with $D^{-1}_{ij}=\Box\delta_{ij}$. For a charged particle with incoming momentum $\textbf{p}$, scattering from the Coulomb field, the transition amplitude between the initial state $|i\rangle=|\textbf{p},0\rangle$ and final state $|f\rangle=|\textbf{p},0\rangle$, with no photon in both states, is given by
{\setlength\arraycolsep{2pt}
\begin{eqnarray}\label{1}
\mathcal A_{\scriptsize{\textbf{p}\textbf{p}}}&=&\langle\textbf{p},0|\int_{\scriptsize{\textbf{x}^{\prime\prime}}} ^{\scriptsize{\textbf{x}^{\prime}}}D\textbf{x}\int D\textbf{A} \,e^{iS_0[\scriptsize\textbf{x}]+iS_{\tiny\textrm{coul}}[\scriptsize\textbf{x}]
+iS_{\tiny\textrm{int}}[\scriptsize\textbf{A},\scriptsize\dot{\textbf{x}}]
+iS_{\tiny\textrm{rad}}[\scriptsize\textbf{A}]}|\textbf{p},0\rangle.
\end{eqnarray}}
with $\textbf{x}(0)=\textbf{x}^{\prime\prime} $ and  $\textbf{x}(T)=\textbf{x}^{\prime}$. Integrating out the gauge field, gives rise to the effective action $S_{\textrm{eff}}$ via
\begin{equation}\label{1}
\langle 0|\int D\textbf{A}e^{iS_{\tiny\textrm{int}}[\scriptsize\textbf{A},\scriptsize\dot{\textbf{x}}]
+iS_{\tiny\textrm{rad}}[\scriptsize\textbf{A}]}|0\rangle=e^{-i\epsilon_0 T}e^{iS_{\tiny\textrm{eff}}[\scriptsize\textbf{J}]},
\end{equation}
where $\epsilon_0$ denotes the photon vacuum energy and
{\setlength\arraycolsep{2pt}
\begin{eqnarray}\label{2}
S_{\scriptsize\textrm{eff}}&=&-2\pi\int d^4x''\int d^4x'\, {J^i}({x}^{\prime\prime})D_{ij}({x}''-{x}'){J^i}({x}'),\\\nonumber
&=&-2\pi\int\frac{d^3k}{(2\pi)^3}\int_0^T dt{''}\int_0^T dt'\,j^i_{\scriptsize\textbf{k}}(t{''})D_{ij,\scriptsize\textbf{k}}(t{''}-t')\bar{j}^j_{\scriptsize\textbf{k}}(t'),
\end{eqnarray}}
where
\begin{equation}\label{1}
D_{ij,\scriptsize\textbf{k}}(t''-t')=\sum_{\lambda=1}^2\int_{-\infty}^\infty\frac{d\omega}{2\pi}\frac{e^{i\omega(t''-t')}}{\omega^2-\textbf{k}^2+i\epsilon}
\epsilon_{i,\lambda\scriptsize\textbf{k}}\overline{\epsilon}_{j,\scriptsize\lambda\textbf{k}}.
\end{equation}
The source $J_i$ and its Fourier transform have the form
{\setlength\arraycolsep{2pt}
\begin{eqnarray}\label{1}
J_i(x')&=&J_i(0,\textbf{x}')=e\dot{x}'_i\,\delta(\textbf{x}'-\textbf{x}_1),\\
J_i(x'')&=&J_i(0,\textbf{x}'')=e\dot{x}''_i\,\delta(\textbf{x}''-\textbf{x}_2),\\
j_{i,\scriptsize\textbf{k}}(t)&=&e\dot{x}_i(t)e^{-i{\scriptsize\textbf{k}\cdot\textbf{x}}}.
\end{eqnarray}}
noting that $\textbf{x}_1=\textbf{x}(t')$ and $\textbf{x}_2=\textbf{x}(t'')$. So, the transition amplitude takes the form
{\setlength\arraycolsep{2pt}
\begin{eqnarray}\label{1}
\mathcal{A}_{\scriptsize{\textbf{p}\textbf{p}}}[\textbf{J}]&=&e^{-i\epsilon_0 T}\langle\textbf{p}|\int_{\scriptsize{\textbf{x}^{\prime\prime}}} ^{\scriptsize{\textbf{x}^{\prime}}}D\textbf{x} \,e^{iS_0[\scriptsize\textbf{x}]+iS_{\tiny\textrm{coul}}[\scriptsize\textbf{x}]+iS_{\tiny\textrm{eff}}[\scriptsize\textbf{J}]}
|\textbf{p}\rangle,\\\nonumber
&=&e^{-i\epsilon_0 T}\int d^3{x}'\int d^3{x}''\bar{\phi}_{\scriptsize\textbf{p}}(\textbf{x}')\langle\textbf{x}',T|\,\textbf{x}^{\prime\prime},0\rangle_J
{\phi}_{\scriptsize\textbf{p}}(\textbf{x}^{\prime\prime}).
\end{eqnarray}}
By substituting (19) in (18) and performing the integration over $\omega$, one obtains [5-7]
{\setlength\arraycolsep{2pt}
\begin{eqnarray}\label{2}
S_{\scriptsize\textrm{eff}}=i\pi \int\frac{d^3k}{(2\pi)^3k}\int_0^T dt_2\int_0^T dt_1\,{e^{-ik|t_2-t_1|}}{j}^i_{\scriptsize\textbf{k}}(t_2)P_{ij,\scriptsize\textbf{k}}\bar{j}^j_{\scriptsize\textbf{k}}(t_1).
\end{eqnarray}}
Now, we define the photon vacuum to vacuum transition amplitude as
\begin{equation}\label{1}
\langle0|0\rangle_{J}=\frac{\mathcal{A}_{\scriptsize{\textbf{p}\textbf{p}}}[\textbf{J}]}{\mathcal{A}_{\scriptsize{\textbf{p}\textbf{p}}}[0]}=e^{i\langle S_{\tiny\textrm{eff}}\rangle_0}
\end{equation}
where
{\setlength\arraycolsep{2pt}
\begin{eqnarray}\label{1}
\langle S_{\scriptsize\textrm{eff}}\rangle_0&=& i\pi\int\frac{d^3k}{(2\pi)^3k}P_{ij,\scriptsize\textbf{k}}\int_0^T dt_2\int_0^T dt_1e^{-ik|t_2-t_1|}\\\nonumber
&\times&\int_{\scriptsize\textbf{x}',\scriptsize\textbf{x}'',\scriptsize\textbf{x}_1,
\scriptsize\textbf{x}_2}{ \bar\phi}_{\scriptsize{\textbf{p}}}(\textbf{x}')\langle\textbf{x}',0|\,\textbf{x}_2,t_2\rangle_0 j^i_{\scriptsize\textbf{k}}(t_2)\langle\textbf{x}_2,t_2|\,\textbf{x}_1,t_1\rangle_0 \bar{j}^j_{\scriptsize\textbf{k}}(t_1) \langle\textbf{x}_1,t_1|\,\textbf{x}^{\prime\prime},0\rangle_0 {\phi}_{\scriptsize{\textbf{p}}}(\textbf{x}'')\\\nonumber
&=&i\pi\int\frac{d^3k}{(2\pi)^3k}P_{ij,\scriptsize\textbf{k}}\int_0^T dt_2\int_0^T dt_1e^{-ik|t_2-t_1|}
\int_{\scriptsize\textbf{x}_1,
\scriptsize\textbf{x}_2}{ \bar\phi}_{\scriptsize{\textbf{p}}}(\textbf{x}_2) j^i_{\scriptsize\textbf{k}}(t_2)\langle\textbf{x}_2,t_2|\,\textbf{x}_1,t_1\rangle_0 \bar{j}^j_{\scriptsize\textbf{k}}(t_1){\phi}_{\scriptsize{\textbf{p}}}(\textbf{x}_1)
\end{eqnarray}}
and
{\setlength\arraycolsep{2pt}
\begin{eqnarray}
\langle\textbf{x}',t'|\,\textbf{x}^{\prime\prime},t''\rangle_0&=&\int D\textbf{x} \,e^{iS_0[\scriptsize\textbf{x}]+iS_{\tiny\textrm{coul}}[\scriptsize\textbf{x}]},\\\nonumber
&=&\int\frac{d^3p}{(2\pi)^3}{\phi}_{\scriptsize\textbf{p}}(\textbf{x}')\overline\phi_{\scriptsize\textbf{p}}(\textbf{x}'')e^{-iE_p(t'-t'')}
\end{eqnarray}}
The second line of (27) represents the spectral expansion of the propagator in terms of the Hamiltonian eigen-function (c.f. (4)). The abbreviation
\begin{equation}\label{1}
\int_{\scriptsize\textbf{x}\ldots}\equiv\int d^3{x}\ldots
\end{equation}
is assumed in (26). Since $|\langle0|0\rangle_{J}|^2$ is the vacuum persistence amplitude, from
\begin{equation}\label{1}
1-|\langle0|0\rangle_{J}|^2=1-e^{-2\textrm{Im}\langle S_{\scriptsize\textrm{eff}}\rangle_0}\simeq2\textrm{Im}\langle S_{\scriptsize\textrm{eff}}\rangle_0,
\end{equation}
one gets the vacuum decay rate, or equivalently, the photon production rate due to the process $|\textbf{p},0\rangle\rightarrow|\textbf{p}',\textbf{k}\rangle$ as
\begin{equation}\label{1}
R=\frac{2}{T}\textrm{Im}\langle S_{\scriptsize\textrm{eff}}\rangle_0,
\end{equation}
On using (27), the second line of (26) becomes
\begin{equation}\label{1}
\int\frac{d^3p'}{(2\pi)^3}\bigg(\int_{\scriptsize\textbf{x}_2} \bar{\phi}_{\scriptsize{\textbf{p}}}(\textbf{x}_2)j^i_{\scriptsize\textbf{k}}(t_2){\phi}_{\scriptsize{\textbf{p}'}}(\textbf{x}_2)
\int_{\scriptsize\textbf{x}_1} \bar{\phi}_{\scriptsize{\textbf{p}'}}(\textbf{x}_1)j^j_{\scriptsize\textbf{k}}(t_1){\phi}_{\scriptsize{\textbf{p}}}(\textbf{x}_1)\bigg)
e^{i(E_p-E_{p'})(t_2-t_1)},
\end{equation}
yielding ($ t_1<t_2$)
{\setlength\arraycolsep{2pt}
\begin{eqnarray}\label{1}
\langle S_{\scriptsize\textrm{eff}}\rangle_0
&=&i\pi\int\frac{d^3k}{(2\pi)^3k}P_{ij,\scriptsize\textbf{k}}\int\frac{d^3p'}{(2\pi)^3}\int_0^T dt_2\int_0^T dt_1e^{i(E_p-E_{p'}-k)(t_2-t_1)}\langle \textbf{p}|j^i_{\scriptsize\textbf{k}}|\textbf{p}'\rangle\langle \textbf{p}'|j^j_{\scriptsize\textbf{k}}|\textbf{p}\rangle.
\end{eqnarray}}
Note that we have excluded the term appearing on setting $t_2=t_1$ in (32) as it has no imaginary part. Provided that $\frac{T}{\lambda}\ll 1$, which holds in $T\rightarrow\infty$ limit, one achieves [5]
{\setlength\arraycolsep{2pt}
\begin{eqnarray}\label{1}
\int^T_0 dt_2\int^T_0 dt_1 e^{i(\lambda+i\epsilon) |t_2-t_1|}&=&2\int^T_0 dt_2\int^{t_2}_0 dt_1 e^{i(\lambda+i\epsilon) (t_2-t_1)},\qquad t_1<t_2\\\nonumber
&=&\frac{2iT}{\lambda+i\epsilon}.
\end{eqnarray}}
The dipole approximation implies $j^i_{\scriptsize\textbf{k}}\simeq e\dot{x}^i$. Accordingly, we obtain
\begin{equation}\label{1}
\langle S_{\scriptsize\textrm{eff}}\rangle_0=-\frac{2e^2}{3\pi}T\int_0^\infty{dk}k\int\frac{d^3p'}{(2\pi)^3}\frac{|\dot{\textbf{x}}_{\scriptsize{\textbf{p}'\textbf{p}}}|^2}
{E_p-E_{p'}-k+i\epsilon}.
\end{equation}
where again we have gained (8). The imaginary part of (34) can be extracted by means of
\begin{equation}\label{1}
\textrm{Im}\frac{f(x)}{x-x_0-i\epsilon}=\pi f(x)\delta(x-x_0).
\end{equation}
So, from (30) one finds the photon emission rate as
{\setlength\arraycolsep{2pt}
\begin{eqnarray}\label{1}
R&=&\frac{4e^2}{3}\int\frac{d^3p'}{(2\pi)^3}\omega_{pp'}|\dot{\textbf{x}}_{\scriptsize{\textbf{p}'\textbf{p}}}|^2,\\\nonumber
&=&\frac{4e^2}{3}\int\frac{d^3p'}{(2\pi)^3}\omega_{pp'}^3|\textbf{x}_{\scriptsize{\textbf{p}'\textbf{p}}}|^2.
\end{eqnarray}}
Multiplying by the photon energy and dividing by the incoming flux and using the classical equation of motion, the above equation yields
{\setlength\arraycolsep{2pt}
\begin{eqnarray}\label{1}
P&=&\frac{4e^2}{3}\int\frac{d^3p'}{(2\pi)^3v}\omega_{pp'}^4|\textbf{x}_{\scriptsize{\textbf{p}'\textbf{p}}}|^2\\\nonumber
&=&\frac{4e^2}{3}\Big(\frac{Ze^2}{2m}\Big)^2\int\frac{d^3p'}{(2\pi)^3v}\,|\langle\textbf{p}'|\nabla\frac{1}{|\bf x|}|\textbf{p}\rangle|^2.
\end{eqnarray}}
So, we come across with the same result obtained by applying the first order perturbation theory.
\section{Born Approximation}
In Born approximation, one replaces the exact wave function with the plane wave  i.e. $\phi_{\scriptsize\textbf{p}}\approx e^{ i\scriptsize{\textbf{p}\cdot\textbf{x}}} $ [1, 2]. Thus by taking into account
\begin{equation}\label{1}
\langle\textbf{p}'|\nabla\frac{1}{|\bf x|}|\textbf{p}\rangle=\int d^3x e^{\scriptsize{-i\textbf{p}'\cdot\textbf{x}}}\nabla\frac{1}{|\bf x|}\,e^{i \scriptsize{\textbf{p}\cdot\textbf{x}}}=4\pi i\,\frac{\textbf{p}'-\textbf{p}}{|\textbf{p}'-\textbf{p}|^2},
\end{equation}
one arrives at
\begin{equation}\label{1}
|\ddot{\textbf{x}}_{\scriptsize{\textbf{p}'\textbf{p}}}|^2=\Big(\frac{Ze^2}{2m}\Big)^2\Big(\frac{4\pi}{m}\Big)^2\frac{1}{|\textbf{v}'-\textbf{v}|^2},
\end{equation}
which yields [1, 2]
\begin{equation}\label{1}
P=\frac{4}{3}\frac{Z^2e^6}{mv^2}\int dv'v'\log\frac{v+v'}{v-v'}
\end{equation}
We shall not evaluate the above integral here. It is plagued with the infinity indicating the infrared divergence caused by the pure Coulomb field. Though, it can be evaluated, for example, to calculate the radiative energy loss of an electron passing through a medium where the screening of the Coulomb potential by the atomic levels renders equation (40) finite [2].

\end{document}